\documentclass{ws-procs9x6}

\newcommand{\bea}{\begin{eqnarray}}
\newcommand{\eea}{\end{eqnarray}}
\newcommand{\beq}{\begin{equation}}
\newcommand{\eeq}{\end{equation}}

\begin{document}

\title{ENHANCED QUANTIZATION:\\ THE PARTICLE ON THE CIRCLE\footnote{Contribution to the XXIXth International Colloquium on Group-Theoretical Methods in Physics
held at the Chern Institute of Mathematics, Nankai, China, August 20-26, 2012.}
}

\author{J. BEN GELOUN}

\address{Perimeter Institute, 31 Caroline Str N, Waterloo, ON N2L 2Y5,
 Canada\\
E-mail: jbengeloun@perimeterinstitute.ca}

\begin{abstract}
Enhanced quantization is an improved program for
overcoming  difficulties which may arise during an ordinary
canonical quantization procedure. We review here how this program applies for a particle on circle.
\end{abstract}


\bodymatter

\section{Introduction}

Conventional canonical quantization
works very well for many systems but it has also  notorious drawbacks.
Enhanced Quantization \cite{Klauder:2012vh}
is an improved quantization procedure  
which has matured through years.\cite{Klauderbook}  It is only recently
that the name of ``Enhanced Quantization'' (EQ) has been found.
This procedure yields a new interpretation of the very process of quantization that encompasses the usual  
canonical formalism and offers additional features as well.

EQ yields a new sense to and improves major points 
which have been  discussed for a long time
 in the ordinary quantization version. For instance,\cite{Klauder:2012uf} 
a) the invariance of the theory under canonical transformations
is ensured under EQ; b) EQ may remove singularities in classical solutions
(Hydrogen atom and quantum toy models \cite{Klauder:2012uf}
and other simple cosmological models  \cite{Fanuel:2012ij});
c) the triviality of certain quantum field models
is now traded for a non trivial behavior (pseudo-free theories)
and d) EQ preserves the metric positivity in quantum gravity kinematics. 

Let us review now the main ingredients of the EQ program.\cite{Klauder:2012uf}
\begin{enumerate}
\item[(i)] From pairs $P$ and $Q$ of self-adjoint operators
(a stronger condition than Hermiticity), one generates
unitary operators acting on a fiducial state $|\eta\rangle$. 
This provides a set of coherent states
\beq
|p,q\rangle  =  e^{-\frac{i}{\hbar}qP} e^{\frac{i}{\hbar}pQ}|\eta\rangle
\eeq
 spanning the Hilbert space $\mathfrak{H}$.

\item[(ii)] Referring to the action principle, arbitrary variations of  $|  \psi(t)\rangle$
for a microscopic system are not accessible to a macroscopic observer
who can only change the velocity or position of the microscopic
system. Hence, the only state she/he could make
are those represented by $|p(t),q(t)\rangle$. 
Thus, restricting the quantum action functional
\beq
A_Q = \int_0^T \, \langle \psi(t)|\, \big[ i\hbar\partial_t - \mathcal{H}\big]
\,|  \psi(t)\rangle\,  dt
\eeq
to only coherent states yields 
\bea
A_{Q(R)}&&= \int_0^T \, \langle p(t),q(t)|\, \big[ i\hbar\partial_t - \mathcal{H}\big]
\,|  p(t),q(t)\rangle \, dt\crcr
&&= \int_0^T \, \Big[ p(t)\dot{q}(t)-H(p(t),q(t))\Big] dt
\label{restact}
\eea
The restricted action \eqref{restact} may be called 
enhanced classical action since is of a classical form 
but, because still $\hbar >0$, it includes certain quantum modifications.  
The usual classical action is given by
\beq
A_C = \int_0^T \, \Big[ p(t)\dot{q}(t)-H_c(p(t),q(t))\Big] dt\,,
\eeq
 where $H_c(p(t),q(t)) = \lim_{\hbar\to 0} H(p(t),q(t))$.
\end{enumerate}
According to this point of view, enhanced classical theory forms a subset of quantum
theory, and they both co-exist just like they do in the real world where
$\hbar >0$ \cite{Klauder:2012vh}.

In this paper, we review the EQ program for 
a classical particle moving on a circle of finite radius.\cite{Geloun:2012bb} 
The particle motion on the circle has been 
studied in different ways\cite{Govaerts:1999ep,Govaerts:2006uu,ChagasFilho:2008ud,
al,Gazeau:2009zz,Kowalski:1998hx,Gonzalez:1998kj,chad} but none of  previous contributions addresses the issue of the relationship between classical and quantum actions which is our main concern here.
The present study leads to yet another set of coherent states which
serves to unite the classical and quantum theories for such a system.
At the quantum level, we find that the non trivial topology induces a possibly shifted momentum easily reabsorbed by a canonical transformation.

\section{Enhanced quantization  on the circle}

Consider a particle on the circle $S^1$ 
parametrized by $\theta \in [-\pi,\pi)$.
At the quantum level, position and momentum operators
$Q$ and $P$, respectively,  obey the commutation relation:
\beq
[Q,P] = i \hbar\,.
\label{heis}
\eeq
The spectrum of the operator $Q$ is bounded in $[-\pi,\pi)$
modulo $2\pi$. The EQ program starts with two self-adjoint
operators $P$ and $Q$. Let us investigate the self-adjoint possibilities
of the operator $P$. 

\noindent{\it Self-adjoint extension of $P$.}
Let us review,\cite{Reed:1975uy} the properties of
$P=-i\hbar \partial_\theta$ acting on $L^2([-\pi,\pi), d\theta)$.

Call   $D(P)$ the domain of $P$.
Consider the inner product for any two functions $\psi,\varphi,\psi',\varphi' \in  L^2([-\pi,\pi), d\theta)$. If we adopt the boundary conditions $\varphi(\pm \pi)= 0$ and make no restriction on $\psi$,
then $D(P) = \big\{\varphi;\; \varphi, \varphi' \in L^2([-\pi,\pi), d\theta); \;\;\varphi(\pi) = \varphi(-\pi) =0 \big\}$ and  $(\psi, P \varphi)=  (P\psi, \varphi)$  holds on $D(P)$. Hence $P$ is symmetric, i.e. $P^\dag = P$ on $D(P)$. However, the domain of $P^\dag$ is larger, 
$D(P^\dag) = \big\{\varphi;\; \varphi, \varphi' \in L^2([-\pi,\pi), d\theta) \big\} \supset D(P)$.

Imposing the boundary condition $\varphi(\pi) = e^{2\pi i\alpha} \varphi(-\pi)$,
for a given $\alpha\in [0,1)$, enlarges the domain of $P$
and reduces the domain of $P^\dag$ so that
\beq
\widetilde{D}(P_\alpha) = \Big\{\varphi;\; \varphi, \varphi' \in L^2([-\pi,\pi), d\theta); \;
\varphi(\pi) = e^{2\pi i\alpha}\varphi(-\pi)\Big\} = \widetilde{D}(P_\alpha^\dag)\,.
\eeq

\noindent{\it Coherent states.}
One denotes $P$ by $P_\alpha$, where $\alpha\in[0,1)$ labels the
different inequivalent representations of the momentum
operator. We work in units such that
$Q$ is dimensionless
and so the dimension of $P$ is that of $\hbar$.
Define eigenvectors $|\theta\rangle $ for
the operator $Q$, obeying
$\langle \theta|\theta'\rangle=\delta_{S^1}(\theta-\theta')$,
where $\delta_{S^1}$ is periodic on $S^1$,
as well as eigenvectors $|n,\alpha \rangle$ of $P_\alpha$, satisfying $\langle n,\alpha| m,\alpha\rangle=\delta_{n,m}$,
 such that
\beq
Q |\theta\rangle  = \theta|\theta\rangle  \,,\quad 
P_\alpha |n,\alpha \rangle = p_{n,\alpha}  |n,\alpha \rangle \,.
\eeq
One has $\langle   \theta | P_\alpha |n,\alpha \rangle = (-i\hbar) \partial_{\theta}
\langle   \theta |n,\alpha \rangle  = p_{n,\alpha} \langle   \theta |n,\alpha \rangle.$
The spectrum of $P_\alpha$ on the circle is such that
$p_{n,\alpha} =\hbar(n +\alpha)$, $(n,\alpha) \in \mathbb{Z}\times [0,1)$ and corresponds to the normalized wave functions  
\beq
\langle   \theta |n,\alpha \rangle  = \frac{1}{\sqrt{2\pi}} e^{i (n + \alpha )\theta} \;.
\eeq
The unitary operators $e^{-\frac{i}{\hbar} q P_\alpha}$ and $e^{-\frac{i}{\hbar} p Q}$, where $(q,p)\in S^1 \times \mathbb{R}$ allow us to 
define a set of states
\bea
|p,q\rangle  = e^{-\frac{i}{\hbar} q P_\alpha}e^{\frac{i}{\hbar} p Q}\;
 |\eta_\alpha\rangle\,,
\label{cohs}
\eea
where $|\eta_\alpha\rangle$ is called the fiducial state. 
One proves that the states \eqref{cohs} are normalized
$\langle  p,q|p,q\rangle  =\langle  \eta_\alpha |\eta_\alpha\rangle=1$,
and they satisfy a resolution of unity:
\beq
\int_{\mathbb{R}\times S^1}  |p,q\rangle \langle  p,q|   \,\frac{dp dq}{2\pi\hbar}
 = I_{\mathfrak{H}}\,.
\label{resolu}
\eeq
The set of states $\{|p,q\rangle\}$ forms an overcomplete
family of normalized states which can be called coherent states.

We can now discuss the dynamics associated with such states
by introducing a general quantum Hamiltonian of the form
${\mathcal H}(P, e^{i Q}, e^{-i Q})$.
Consider the restricted quantum action associated with
$|\psi(t)\rangle\rightarrow|p(t),q(t)\rangle$ which leads to
\beq
A_{Q(R)} = \int_0^T \;
\langle p(t),q(t)| \Big[i\hbar \partial_t\; - \;{\mathcal H} \Big]
| p(t),q(t) \rangle\; dt\,.
\eeq
A class of fiducial vectors  $|\eta_\alpha\rangle$ is choosen 
in the domain of both $Q$ and $P_\alpha$  such that these states satisfy
\beq
\langle \eta_\alpha |Q |\eta_\alpha\rangle =0
\qquad \text{and} \qquad
\langle\eta_\alpha |P_\alpha |\eta_\alpha\rangle =\hbar \alpha\,.
\label{qpmean}
\eeq
Hence $|\eta_\alpha\rangle$ obeying the condition  \eqref{qpmean}
can be considered analogs of the ``physically centered'' 
fiducial vectors $|\eta\rangle$ for the canonical case.

A direct evaluation  using \eqref{qpmean} yields
\beq
\langle p(t),q(t)| \Big[i\hbar \partial_t \Big] | p(t),q(t) \rangle
= (\hbar\alpha + p)  \dot{q} \,.
\label{palph}
\eeq
Moreover, one proves that
\beq
H_\alpha(p(t),q(t)) =\langle\eta_\alpha |   {\mathcal H} (P_\alpha + p, e^{i (Q+q)}, e^{-i (Q+q)})
| \eta_\alpha \rangle \,.
\label{halp}
\eeq
Equations \eqref{palph} and \eqref{halp} imply that the restricted quantum action is of the form
\beq
A_{Q(R)} = \int_0^T\;
\Big\{
  \Big[  p  \dot{q} -
 H_\alpha(p(t),q(t))\Big]+\hbar\,\alpha \dot{q}\Big\}dt\,.
\eeq
One notices that the part
$\widetilde{A}_{C} = \int_0^T[ p\dot{q}    - H_\alpha(p(t),q(t))  ]dt$,
 as in the ordinary situation \cite{Klauder:2012vh} can be related to
a classical action $A_C$ up to $\hbar$ corrections using 
\beq
 H_\alpha(p,q)  = H_{c,\alpha}(p,q) +O(\hbar;p,q)\,.
\eeq
$H_{c,\alpha}(p,q)$ is viewed as the usual classical Hamiltonian.
Note that the quantum parameter $\alpha$
induces a surface term $\hbar \alpha \dot{q} $ in $A_{Q(R)}$ which  makes no influence on the enhanced classical equations of motion whatsoever.  Thus, one has
\bea
A_{Q(R)} = A_C + O(\hbar)\;.
\eea
Let us apply the above formalism to the particular instance 
of a particle governed by the following (periodic) dynamics
\bea
&&
\mathcal{H} (P_\alpha,e^{iQ}, e^{-iQ}) = P_\alpha^2 + V(e^{iQ}, e^{-iQ})\,,
\label{hamilsin}\\
&&
 V(e^{iQ}, e^{-iQ})
= a_0+\sum_{n=1}^{m} \big[ a_n \cos n Q + b_n \sin n Q\big]\,,
\nonumber
\eea
in mass units such that $1/2\mu=1$ and with 
 $m$ a positive integer. As a  fiducial vector, we consider
\bea
&&
\eta_\alpha(\theta)  := \langle \theta | \eta_\alpha\rangle =N e^{  (r/\hbar) (\cos\theta - 1)  + i \alpha \theta }\,,
\crcr
&&
N = \Big[ 2\pi e^{-2r/h}\, I_0\Big( \frac{2r}{\hbar}\Big) \Big]^{-1/2},
\label{fidu1}
\eea
  where $r/\hbar>0$, $I_0(z)$ stands for a
modified Bessel funtion\cite{abram}, and $N$ is a normalization  fixed 
by $\langle \eta_\alpha | \eta_\alpha\rangle=1$. 
Among its properties, $|\eta_\alpha(\theta)|$ is even, 
periodic and fulfills \eqref{qpmean}.
For large $r/\hbar\gg1$ and for $|\theta|\le\pi$, 
one makes the approximation
\beq
|\eta_\alpha(\theta)|^2 = N^2
e^{\frac{2r}{\hbar}  [\cos\theta-1 ]}
\lesssim  K N^2
e^{-\frac{r}{\hbar}\theta^2}\,,
\label{spa}
\eeq
for a large constant $K$ impying that
$|\theta|\lesssim \sqrt{\hbar/r}$ is small.
Therefore, $\eta_\alpha(\theta)$ acts as a large
$\theta$-value cut-off.
Calculating the diagonal coherent state matrix elements of $\mathcal{H}$, and denoting $\alpha'=\hbar\alpha$, we come to the restricted
quantum action given by
\beq
A_{Q(R)} = \int_0^T \; \Big\{ (p + \alpha')\dot{q}
-\Big[  (p+\alpha')^2
+  V(e^{iq}, e^{-iq})\Big]  +O(\hbar)
\Big\}dt\,.
\label{qar}
\eeq
Therefore, up to constants and  a canonical shift in momentum ($p\rightarrow p  - \alpha'$),
\bea
A_{Q(R)} &=& \int_0^T \; \Big\{ p\dot{q}
-\Big[ p^2  +  V(e^{iq}, e^{-iq}) \Big]  +O(\hbar) \Big\}dt \crcr
&=& A_{C}
+ O(\hbar)\,.
\label{aqr}
\eea
Evaluating \eqref{qar}, 
some expectation values (for instance $\langle\eta_\alpha |  (P_\alpha + p)^2| \eta_\alpha \rangle$) contain  terms $\alpha'$ of order $O(\hbar)$.
We have preferred to remove these terms in a unified way by
making a canonical shift of momenta at the last stage \eqref{aqr}.
Another precision,  the quantity
$\langle\eta_\alpha | \;V(e^{i (Q+q)}, e^{-i (Q+q)}) \;| \eta_\alpha \rangle$
contains, strictly speaking, at the first order of approximation,
terms of order $O(\hbar/r)$. These are subtleties without any consequence for the result.

\section*{Acknowledgements} 
The organizers of the XXIXth International Colloquium on Group-Theoretical Methods in Physics, Nankai, China, are warmly thank for their welcome and hospitality. Discussions with John R. Klauder are gratefully
aknowledged. 
Research at Perimeter Institute is supported by the Government of Canada through
Industry Canada and by the Province of Ontario through the Ministry of Research and Innovation.

\end{document}